\documentclass[12pt]{article}

\usepackage{scicite}
\usepackage{times}
\usepackage{xcolor}
\usepackage{upgreek}
\usepackage{siunitx}
\usepackage{amsmath}
\usepackage[utf8]{inputenc}
\usepackage{graphicx}
\usepackage{ulem}

\newenvironment{sciabstract}{%
\begin{quote} \bf}
{\end{quote}}

\title{Experimental detection of vortices in magic-angle graphene} 

\author {Marta Perego,$^{1\ast}$ Clara Galante Agero,$^{1}$ Alexandra Mestre
Tor\`a,$^{1}$ El\'ias Portol\'es,$^{1}$\\ Artem O.
Denisov,$^{1}$ Takashi Taniguchi,$^{2}$ Kenji Watanabe,$^{3}$\\ Filippo
Gaggioli,$^{4,5}$ Vadim Geshkenbein,$^{4}$ Gianni Blatter,$^{4}$\\ Thomas
Ihn,$^{1,6}$ and Klaus Ensslin$^{1,6}$\\ \\ 
\normalsize{$^{1}$Laboratory for
Solid State Physics, ETH Zurich,~CH-8093~Zurich, Switzerland,}\\
\normalsize{$^{2}$Research Center for Materials Nanoarchitectonics, National Institute for Materials Science}\\
\normalsize{1-1 Namiki, Tsukuba 305-0044, Japan,}\\
\normalsize{$^{3}$Research Center for Electronic and Optical Materials, National Institute for Materials Science,,}\\
\normalsize{1-1 Namiki, Tsukuba 305-0044, Japan,}\\
\normalsize{$^{4}$Institute for Theoretical Physics, ETH Zurich,~CH-8093
Zurich, Switzerland,}\\
\normalsize{$^{5}$Department of Physics, Massachusetts Institute of Technology, Cambridge,~MA-02139, USA,}\\
\normalsize{$^{6}$Quantum Center, ETH Zurich,~CH-8093 Zurich, Switzerland}\\
\normalsize{$^\ast$To whom correspondence should be addressed; E-mail:
mperego@phys.ethz.ch.} }

\date{}

\begin{document} 


\baselineskip24pt


\maketitle 


\begin{sciabstract} 
%
%
%The tunability of superconducting magic-angle twisted-layer graphene films
%elevates this material system to a promising candidate for superconducting
%electronics.  We implement a gate-tuned Josephson junction within a
%magic-angle twisted four-layer graphene film. Field-dependent measurements of
%the critical current show a Fraunhofer-like pattern that differs markedly from the standard pattern with characteristics typical for a weak transverse
%screener. We observe sudden shifts that we associate with vortices jumping
%into and out of the leads and present a quantitative analysis of this
%phenomenon. By tuning the leads to the edge of the superconducting dome, we
%observe fast switching between superconducting and normal states, an effect
%that we associate with thermally activated vortex dynamics. Time-dependent
%measurements of these switching phenomena provide us with the vortex energy
%scale of order Kelvin and an estimate for the London penetration depth of a
%few micrometers, in agreement with recent kinetic inductance measurements on
%twisted graphene films. Our results prove the utility of our junction as a
%sensor for vortex detection, allowing us to extract fundamental properties of
%the 2D superconductor.

The tunability of superconducting magic-angle twisted-layer graphene films elevates this material system to a promising candidate for superconducting electronics. We implement a gate-tuned Josephson junction in a magic-angle twisted four-layer graphene film. Field-dependent measurements of the critical current show a Fraunhofer-like pattern that differs from the standard pattern with characteristics typical for a weak transverse screener. We observe sudden shifts associated with vortices jumping into and out of the leads. By tuning the leads to the edge of the superconducting dome, we observe fast switching between superconducting and normal states, an effect associated with vortex dynamics. Time-dependent measurements provide us with the vortex energy scale and an estimate for the London penetration depth, in agreement with recent kinetic inductance measurements on twisted graphene films. Our results prove the utility of our junction as a sensor for vortex detection, allowing us to extract fundamental properties of the 2D superconductor.

\end{sciabstract}

\section*{Main Text} 

\paragraph*{Introduction}

Twisted-layer graphene has emerged as a new platform for realizing non-trivial
correlated states \cite{cao2018correlated, yankowitz2019tuning,
lu2019superconductors}, with superconducting bi- and multilayer systems
attracting particular interest recently \cite{cao2018unconventional,
park2021tunable, hao2021electric, carr2017twistronics, park2022robust}.
Superconductivity in these two-dimensional (2D) structures can be electrically
tuned, enabling the implementation of gate-defined devices.  Much research has
focused on magic-angle twisted bilayer graphene (MATBG), that has become a
versatile platform for superconducting electronics \cite{Rodan-Legrain2021,
deVries2021, diez2023symmetry, portoles2022tunable}.  Recently,
alternating-twist magic-angle multilayer graphene structures have been
realized as a new family of moiré superconductors \cite{kim2022evidence,
park2022robust, zhang2022promotion, burg2022emergence}.  Superconductivity in
these multilayer structures is characterized by higher critical currents,
critical magnetic fields, and critical temperatures than in MATBG. Moreover,
their band structure can be tuned by a transverse electrical field, the
so-called displacement field \cite{khalaf2019magic}, providing additional
versatility for device operation.  As a result, alternating-twist magic-angle
multilayer graphene structures are promising candidates for future
superconducting electronic devices.

Here, we implement a gate-defined Josephson junction (JJ) in four-layer
twisted graphene (MAT4G). Exposing the junction to a perpendicular magnetic
field $B$, we observe a distinct Fraunhofer-like pattern in the junction
critical current $I_\textnormal{cj}(B)$.  The interference pattern, see
Fig.~1D, differs markedly from the one observed in standard junctions
\cite{Tinkham2004}, a consequence of the extremely weak transverse magnetic
screening power of these ultrathin films: i) The periodicity of the pattern is
given by the flux $\Phi_W = B W^2$ with $W$ the junction width, a factor $W/2
\lambda_\textnormal{L}$ larger than the characteristic flux $\Phi_\lambda = 2
B W \lambda_\textnormal{L}$ in usual junctions, where $\lambda_\textnormal{L}$
denotes the London penetration depth~\cite{Tinkham2004}. And ii), the maxima
in the Fraunhofer-like pattern decay slowly $\propto 1/\sqrt{B}$, rather than
the usual decay $\propto 1/B$; recognizing these differences, in the
following, we refer to our experimentally observed interference pattern as a
Fraunhofer pattern (FP). Interestingly, our Fraunhofer pattern exhibits
pronounced jumps, see Fig.~1, E and F, that we attribute to vortices jumping
into and out of the superconducting leads. Our Josephson junction then serves
as a sensor allowing for an indirect detection of vortices in gated atomically
thin materials, circumventing the obstacles of traditional vortex imaging
techniques \cite{Essmann1967, Hess1989, Bending1999, Embon2015}. Furthermore,
we profit from the well developed phenomenology for these ultrathin materials
\cite{Gurevich1994, MosheKoganMints2008, Clem2010, KoganMints2014,
KoganMints_PC2014, Gaggioli2024} that provides us with a strong basis for the
quantitative analysis of our experiments, independent of the microscopic
origin of superconductivity in this correlated material that is still under
debate.

\paragraph*{Setup}
%Device description  

We have engineered a gate-defined JJ in a MAT4G film of width
$W=\SI{1.1}{\upmu m}$, length $L = 6W$, and thickness $d \approx \SI{1}{nm}$
\cite{rickhaus2020electronic}. The device depicted in Fig.~1A features an
alternating twist angle of $1.64^\circ \pm 0.07^\circ$ and is tuned with three
gates. A graphite bottom gate (BG) and a gold top gate (TG) are used to
control independently the density $n_\textnormal{l}$ and the displacement
field $D_\textnormal{l}$ in the leads. A junction of width $W=\SI{1.1}{\upmu
m}$ and length $L_\textnormal{j}=\SI{150}{nm}$ is defined using a gold finger
gate (FG), as shown in Figs.~1,~A~and~B. The density $n_\textnormal{j}$ and
displacement field $D_\textnormal{j}$ in the junction are tuned using both FG
and BG. Three gates are used to tune the four parameters $n_\textnormal{l}$,
$D_\textnormal{l}$, $n_\textnormal{j}$, and $D_\textnormal{j}$, implying that
one parameter out of four cannot be independently controlled.  All experiments
are carried out at $T = \SI{55}{mK}$ unless stated otherwise. The formation of
a JJ is confirmed by the observation of the d.c. Josephson effect shown in
Fig.~2C, as well as by magnetic interference measurements, with the resulting
Fraunhofer pattern shown in Fig.~1D. Further details of the device fabrication
and gate control are presented in the Supplementary Text (see figs.~S1 and
S2).  Given the nanometer scale thickness $d$ of the film $d \ll
\lambda_\textnormal{L}$, the device belongs to the class of weak transverse
screeners, with the effective screening power given by the Pearl length
$\Lambda = 2 \lambda_{\textnormal{L}}^2/d \ll
\lambda_\textnormal{L}$~\cite{Pearl1964}. With the width $W \ll \Lambda$,
external magnetic fields $H$ penetrate the entire sample and $B \approx H$.

\paragraph*{Bulk superconductivity} We first analyse the bulk
superconductivity observed in our MAT4G device. In Fig.~2A the voltage $V$
measured at constant probe current $I$ in a 4-terminal configuration not
crossing the junction is shown as a function of $n_\textnormal{l}$ and
$D_\textnormal{l}$ (see Supplementary Text fig.~S3B for the same measurement
taken across the junction). Superconductivity is observed at moiré filling
factors $\nu$ between $-3.5 \leq \nu \leq-2$ (hole carriers with $T_c$ up to
$\SI{2}{K}$, see fig.~S3C) as well as $2 \leq \nu \leq 3.5$ ($T_c$ up to
$\SI{1}{K}$), with $\nu = \pm 4$ at full filling, see also Supplementary Text.
Strong out-of-plane displacement fields $|D_\textnormal{l}| / \epsilon_0 \geq
\SI{0.6}{V/nm}$ ($\epsilon_0$ denotes the vacuum permittivity) quench the
superconducting state due to the hybridization of the flat bands with the
dispersive bands \cite{khalaf2019magic, park2022robust, burg2022emergence,
zhang2022promotion}.  The bulk critical current $I_\textnormal{cb}$ is
measured to be larger than $\SI{200}{nA}$ in the superconducting regime for
holes and larger than $\SI{100}{nA}$ for electrons (see figs.~S3, D and E).
Further characterization of the superconducting domes is discussed in the
Supplementary Text.

The dependence of the bulk critical current $I_\textnormal{cb}$ on the
magnetic field $B$ applied normal to the sample plane is shown in Fig.~2B for
the device tuned at the green square depicted in Fig.~2A. The critical current
is seen as a peak in the differential resistance $R=dV/dI$ which is evaluated
numerically from d.c. data (transition between dark blue in Fig.~2B, zero
resistance, to a finite one, light blue), where the device switches to the
resistive state due to vortex motion, see inset in Fig.~2B. The field
dependence $I_\textnormal{cb}(B)$ is governed by surface- and bulk pinning of
vortices\cite{BeanLivingston1964, LarkinOvch1979, Maksimova1998, Plourde2001,
Gaggioli2024}. The linear drop in $I_\textnormal{cb}(B)$ at small fields is
characteristic of the surface barrier preventing vortices from entering the
sample \cite{Maksimova1998, Plourde2001, Gaggioli2024}, as is typical for such
films. By linearly fitting the critical current $I_\textnormal{cb}$ versus
magnetic field $B$, we extract a value for the surface penetration field
$B_\textnormal{s} \approx \SI{100}{mT}$, see orange dashed line in Fig.~2B.
Combining this result with the zero-field critical current
$I_\textnormal{cb}(0) = \SI{230}{nA}$ and using the relation $\partial_B
I_\textnormal{cb}(B) = - d W^2/2 \mu_0 \lambda_\textnormal{L}^2$ derived in
Ref.\ \cite{Gaggioli2024}, we arrive at an estimate for the London penetration
depth $\lambda_\textnormal{L}$ of order $\SI{10}{\upmu m}$, that compares
favorably with values of a few $\upmu$m obtained from other estimates (here,
$\mu_0$ is the vacuum permeability). For example, we may use the bulk critical
current $I_\textnormal{cb}(0) \approx \SI{230}{nA}$ measured in Fig.~2B,
obtain the critical current density $j_\textnormal{cb} \approx
\SI{2.1e4}{A/cm^2}$, and take this value as a lower bound for the depairing
current density $j_0 = \Phi_0 / 3\sqrt{3}\pi\mu_0\lambda_\textnormal{L}^2\xi$
(with $\Phi_0 = h/2e$ denoting the flux unit). Extracting a value
$\xi\approx\SI{40}{nm}$ for the coherence length \cite{park2022robust,
burg2022emergence, zhang2022promotion} from the measured critical field
$B_\textnormal{c2}(T)$ as a function of temperature (see fig.~S4 in the
Supplementary Text), we then find an estimate $\lambda_\textnormal{L} \approx
\SI{3.5}{\upmu m}$. Finally, the low value in the saturation of the critical
current at large fields, see Fig.~2B, is testimony of weak bulk pinning
\cite{Gaggioli2024}.

%Explanation of tuning JJ - Fig. 2C

\paragraph*{Implementing the JJ} Keeping the leads in one of the
superconducting regimes, we form a JJ in our sample by tuning the junction
region into a resistive state.  The high tunability of our device allows us to
define a JJ in several ways by exploring the parameter space of density $n$
and displacement field $D$ in the leads and the junction (see Fig.~2A). In
Fig.~2C, we illustrate the formation of a JJ with a reduced critical current
$I_\textnormal{cj} < I_\textnormal{cb}$ by tuning the junction density
$n_\textnormal{j}$ at fixed $n_\textnormal{l} = \SI{-4.5e12}{cm^{-2}}$ and
$D_\textnormal{j} = 0$ (the displacement field $D_\textnormal{l}$ is slaved
to $n_\textnormal{j}$ and follows the yellow dashed line within the
superconducting dome in Fig.~2A). The differential resistance $R$ then
exhibits two steps, i) at low currents $I_\textnormal{cj}$ when the junction
turns resistive, and ii) at high currents $I_\textnormal{cb}$ when the leads
switch to the resistive state, see orange line-trace in Fig.~2C. The two
critical currents coalesce when the junction density $n_\textnormal{j}$ enters
the superconducting state between the white dashed lines in Fig.~2C, resulting
in a uniform superconducting device SSS. Thus, tuning the junction away from
this region, we can implement a JJ with a desired critical current
$I_\textnormal{cj}$---we refer to this weak-link configuration as SJS, with J
denoting the resistive junction region, notwithstanding its nature,
metallic, semiconducting, or insulating. We note that our MAT4G device admits
$D_\textnormal{j}$ as an additional tuning knob for changing
$I_\textnormal{cj}$ as compared to junctions defined in MATBG
\cite{deVries2021, Rodan-Legrain2021} (see fig.~S5F).

Regarding the issue of reproducibility in the formation of the JJ, we have
performed multiple measurements of $I_\textnormal{cj}$ as a function of
$n_\textnormal{j}$ as well as magnetic field $B$ (see figs.~S6, S7, and S12).
We find that a reproducible JJ is only formed when $n_\textnormal{j}$ is tuned
to the full-filling peaks ($n_\textnormal{j} = \pm \SI{6.2e12}{cm^{-2}}$) as
indicated by the red dots in Fig.~2A. When $n_\textnormal{j}$ is tuned to any
correlated insulator state, the measurements are not reproducible in time.
Such reproducibility is an important feature in all practical use-cases of
superconducting electronics.

%Experimental paragraph on Fraunhofer pattern - Fig. 1D
%
\paragraph*{Junction in magnetic field} A typical magnetic interference
measurement with the device tuned to the junction regime is presented in
Fig.~1D. The current $I$ is swept from negative to positive values while
measuring the voltage drop across the junction, all at a fixed magnetic field
$B$. After each current sweep, the magnetic field $B$ is stepped and a new
trace is recorded. The leads are set to the superconducting state
($n_\textnormal{l} =\SI{-3.5e12}{cm^{-2}}$, ${D_\textnormal{l}/\epsilon_0 =
\SI{0.2}{V/nm}}$, see the light blue rhombus in Fig.~2A) and the junction is
gated to full filling ($n_\textnormal{j} = \SI{-6.2e12}{cm^{-2}}$,
${D_\textnormal{j}/\epsilon_0 = \SI{0.45}{V/nm}}$, see the red dot in
Fig.~2A); we call this the BRB (for blue-red-blue) setting with a
corresponding identifier in the top-right corner of Fig.~1D. Note that a SJS
device is characterized by a pair of points in Fig.~2A specifying density and
displacement fields in the junction and the leads. By applying a perpendicular
field $B$, the critical current of the junction is suppressed and modulated,
resulting in the Fraunhofer interference pattern shown in Fig.~1D---the
observed pattern is typical for a short junction~\cite{barone1982physics} and
exhibits the hallmarks of a weak screening device~\cite{MosheKoganMints2008,
Clem2010}.  We observe field-induced oscillations of the critical current
$I_\textnormal{cj}(B)$ with the interference period $\Delta{B}\approx
\SI{3}{mT}$. The magnetic interference measurements are symmetric in current
and do not show any skewness. Thanks to the high tunability of our device, we
can study these interference patterns throughout the phase space of both
junction and leads (see Supplementary Text for further measurements with
different parameters $n_\textnormal{l}$ and $D_\textnormal{l}$).

To further prove that the measured interference pattern is due to the JJ
rather than sample inhomogeneity, we carry out the same measurement with the
JJ tuned into the superconducting lobe (SSS) as shown in Fig.~2D (with
$n_\textnormal{j} = n_\textnormal{l} \approx \SI{-4.5e12}{cm^{-2}}$ and
$D_\textnormal{j}/\epsilon_0 = D_\textnormal{l}/\epsilon_0 \approx
\SI{0.2}{V/nm}$, see the pink triangle in Fig.~2A). In this measurement, the
critical current $I_\textnormal{cj}$ agrees with the one measured for the bulk
up to small oscillations in $I_\textnormal{cj}(B)$ at low fields; their
periodicity closely matches the one observed in the pattern of Fig.~1D and we
attribute them to a slight mismatch in the tuning between the leads and the
junction regions. No interference pattern is observed when the device is fully
superconducting, compare Fig.~1D with Fig.~2D.

%Theoretical interpretation - fitting of Fig. 1D
%
\paragraph*{Fraunhofer interference pattern} For a JJ under weak screening
conditions with $\Lambda \gg W$  \cite{Rosenthal1991, Humphreys1993,
MosheKoganMints2008, Clem2010} the critical current shows a non-standard
Fraunhofer pattern $I_\textnormal{cj}(B)$.  In a conventional Josephson
junction, the magnetic field in the leads is screened on the distance
$\lambda_\textnormal{L}$, the bulk London penetration depth
\cite{Tinkham2004}. As a consequence, the gauge-invariant phase difference
$\Delta \gamma_B(y) = 2\pi [\Phi_\lambda(B)/\Phi_0] (y/W)$ across the junction
involves the flux $\Phi_\lambda (B) = B (2\lambda_\textnormal{L} +
L_\textnormal{j}) W$. However, when screening is weak, the field penetrates
the leads completely and the corresponding result~\cite{MosheKoganMints2008,
Clem2010}
\begin{equation}\label{eq:dga_at-th-film}
   \Delta \gamma_B(y) \approx 1.7 \frac{\Phi_W(B)}{\Phi_0} \sin(\pi y/W)
\end{equation}
deviates from the standard expression. First, the relevant flux associated
with the junction is now given by the junction width $W$ only, $\Phi_W(B)= B
W^2$, provided that the film leads are longer than the film width, $L \gg W$,
which is the case in our device.  As a consequence, the oscillations in the
Fraunhofer pattern appear on the scale $\Delta B \approx \Phi_0/ W^2$ rather
than the usual periodicity $\Delta B = \Phi_0/ (2\lambda_\textnormal{L} +
L_\textnormal{j}) W$. Hence the field-to-flux conversion involves only the
width $W$.

Second, the usual linear shape $\Delta \gamma_B(y) \propto y/W$ is replaced by
a sine-function, $\Delta \gamma_B(y) \propto \sin (\pi y/W)$. This is again
due to the deep penetration of the field into the film, on the scale $W$
rather than $\lambda_\textnormal{L}$: in thick superconductors, screening
currents flow on the London scale $\lambda_\textnormal{L}$ and turn around
sharply in the device corners on this small length scale. In atomically thin
films, however, screening is weak and the currents bend smoothly around
corners on the scale $W$ of the sample, see Fig.~1C. As a result, the
component of the screening current density $j_{\textnormal{s}y}(x=0^+,y)$
flowing {\it along} the junction edge vanishes on the scale $W$ when
approaching the junction edges at $\pm W/2$, see Fig.\ 2 in Ref.\
\cite{Clem2010}. With the phase difference across the junction given by the
relation \cite{Clem2010} $\partial_y \Delta\gamma_B(y) \propto
j_{\textnormal{s}y}(0^+,y)$, one finds that $\Delta\gamma_B(y)$ flattens at
the junction edges, which explains the $\sin(\pi y/W)$ factor in
\eqref{eq:dga_at-th-film}.  This seemingly minor correction has profound
consequences for the Fraunhofer pattern at large fields $B$, producing a slow
decay of the maxima $\propto 1/\sqrt{B}$ in the pattern instead of the
standard $1/B$ behaviour.

To relate these insights to our experiment, we find the junction critical
current $I_\textnormal{cj}(B)$ by integration over the junction dimensions:
assuming a sinusoidal current--phase relation $j_\textnormal{j}(\Delta \gamma)
= j_\textnormal{cj}\sin (\Delta\gamma + \gamma_0)$ with $\gamma_0$ a free
shift parameter, the integral over $W$ can be evaluated exactly
\cite{MosheKoganMints2008, Clem2010} in terms of the Bessel function $J_0$,
\begin{equation}\label{eq:IcH0}
   \frac{I_\textnormal{cj}(B)}{I_\textnormal{cj}(0)} 
   = \big| J_0[1.7\> \Phi_W(B)/\Phi_0] \big|,
\end{equation}
where the choice $\gamma_0 = \pm\pi/2$ produces the largest, hence critical,
current. The Bessel function $J_0$ then replaces the $\mathrm{sinc}$-function
characterizing the Fraunhofer pattern in the standard context
\cite{Tinkham2004}. The orange line in Fig.~1D is a fit to the data that makes
use of the width $W = \SI{1.1}{\upmu m}$ of the film in the determination of
the flux $\Phi_W(B) = B W^2$. Given that the distance between consecutive
zeros of the Bessel function $J_0(s)$ is $\Delta s \approx 3.1$, we recover
the periodicity $\Delta B \approx (\Delta s/1.7) \Phi_0/W^2 \approx
\SI{3}{mT}$. Note that the zeros in $J_0(s)$ become truly equidistant only at
large values of the magnetic field. Furthermore, the maxima are well tracked
by the slow decay $\propto 1/\sqrt{B}$, see the dashed yellow line at $B < 0$.

The measured interference patterns fit with the theoretical prediction of Eq.\
\eqref{eq:IcH0} both when the JJ is tuned to full filling for electrons
($n_\textnormal{j} = \SI{6.2e12}{cm^{-2}}$ and $n_\textnormal{l} =
\SI{4.5e12}{cm^{-2}}$) and full filling for holes ($n_\textnormal{j} =
\SI{-6.2e12}{cm^{-2}}$ and $n_\textnormal{l} = \SI{-4.5e12}{cm^{-2}}$), see
red dots in Fig.~2A. Our excellent fit of the Fraunhofer pattern goes beyond
the results of similar experiments~\cite{deVries2021, Rodan-Legrain2021,
park2022robust}. The good agreement between the theoretical prediction and the
experimental data holds at fields below $|B| \approx \pm\SI{10}{mT}$, i.e.,
including the first three minima; thereafter, the fit does not work anymore
due to the presence of sharp shifts in the pattern of the size of a fraction
of $\Phi_0$. We attribute these shifts to vortices jumping into and out of the
leads, as further discussed below.

%Observation of jumps in FP - experimental paragraph about Fig. 1 D,E,F
%
\paragraph*{Jumps in the Fraunhofer pattern} The measured interference
patterns show sudden shifts, i.e., jumps in $I_\textnormal{cj}(B)$, see
Figs.~1,~D--F.  The measurement protocol for the Fraunhofer pattern described
above produces interference maps $I_\textnormal{cj}(B)$ over time scales of
hours. The sudden shifts appear in all of our junction tunings, i.e.,
independent of gate voltages, see fig.~S11.

Pronounced jumps in $I_\textnormal{cj}$, see black arrows in Figs.~1,~E and F,
are observed in an electron-doped state (with $n_\textnormal{l} =
\SI{4.2e12}{cm^{-2}}$, $D_\textnormal{l}/\epsilon_0 = \SI{-0.3}{V/nm}$ and
$n_\textnormal{j} = \SI{6.2e12}{cm^{-2}}$, $D_\textnormal{j}/\epsilon_0 =
\SI{-0.5}{V/nm}$, see the orange rhombus and red dot in Fig.~2A, we name it
the orange-red-orange or ORO setting).  They are present in both increasing
and decreasing sweeps of the magnetic field, are symmetric in the current, and
are stable over typical time intervals of order $10^3$ seconds, see fig.~S14.
Some jumps are found to be very reproducible as they tend to occur at similar
values of the magnetic field, see the ones highlighted by the green arrows in
Figs.~1,~E~and~F at $\pm\SI{5.8}{mT}$ and the discussion of Fig.~3 below.

To further study these shifts, we measure $I_\textnormal{cj}(B)$ in a {\it
consecutive} manner, alternating the direction of the magnetic field-sweep
in-between measurements and increasing the magnetic field range after each
reversal, see Fig.~3 and fig.~S13 (we use the same ORO tuning as for
Figs.~1,~E~and~F).  The results exhibit some degree of hysteresis, as the
central peak (see grey dashed line) of the interference pattern is shifted
between sweeps.  Reproducible jumps are observed, see orange dashed lines at
$\pm\SI{5.8}{mT}$, the jumps marked by green arrows coincide with the events
observed in Figs.~1,~E~and~F. Furthermore, the results tend to be symmetric
for decreasing/increasing magnetic fields, see black traces in Fig.~3 measured
for a decreasing field {\it magnitude} (see fig.~S9 for the extraction method
of these traces).  The range of magnetic-field sweeps influences the
appearance of shifts, with the first few curves in Fig.~3 at small sweep-range
exhibiting almost no jumps.

%Theoretical interpretation of jumps - to be completed by theorists
%
\paragraph*{Vortex jumps} With these shifts present in all of our device
configurations, we rule out a possible origin related to the experimental
setup (see Supplementary Text and fig.~S10). We rather attribute these sudden
shifts to vortices jumping in and out of the superconducting leads in the
vicinity (closer than $2W$) of the junction \cite{KoganMints2014}.
Similar observations have been reported in the past: the influence of
vortices on the FP of cross-strip Josephson junctions has been discussed for
Pb-Bi films in Refs.~\cite{Hyun1987, Hyun1989} and pronounced steps similar to
those reported here have been observed in step-edge junctions of a cuprate
superconductor \cite{Mitchell1999b}.  The deliberate placing of a vortex near
a Josephson junction has been studied in Ref.\ \cite{Golod2010}, with the
underlying theory developed later by Kogan and Mints~\cite{KoganMints2014,
KoganMints_PC2014}.

The presence of a vortex (at the position $\mathbf{R}_\textnormal{v} =
(x_\textnormal{v},y_\textnormal{v})$ and with a flux parallel to the field)
changes the phase pattern at the junction, $\Delta \gamma(y) \to \Delta
\gamma(y; \mathbf{R}_\textnormal{v}) = \Delta \gamma_B(y) + \Delta
\gamma_\textnormal{v}(y;\mathbf{R}_\textnormal{v})$, adding a step-like
contribution $|\Delta \gamma_\textnormal{v}(y;\mathbf{R}_\textnormal{v})| <
\pi$, see figs.~S15--S17 in the Supplementary Text. The phase $|\Delta
\gamma_\textnormal{v}(y;\mathbf{R}_\textnormal{v})|$ as a function of $y$ (see
Supplementary Text for an explicit expression) smoothens and decreases in
amplitude with increasing distance $x_\textnormal{v}$ of the vortex from the
junction. Since the vortex currents (${j_\textnormal{v}}$, blue in Fig.~1C)
flow opposite to the screening currents (${j_\textnormal{s}}$, green in
Fig.~1C), the presence of a vortex {\it decreases} the effect of the field $B$
and the Fraunhofer pattern shifts to the right upon vortex entry at positive
fields $B>0$ (the shift is to the left for a vortex with opposite circularity
entering at $B < 0$). This is exactly what is seen in the experiment. For
example, in Figs.~1, G and H, we fit the experimental traces of Figs.~1, E and
F by having 7 (in Fig.~1G) and 4 (in Fig.~1H) vortices of $\pm$ polarity enter
(upon increasing $|B|$) or leave (upon decreasing $|B|$) the device, see black
and green arrows.  In our fit, we place the vortices in the film center,
i.e., $y_\textnormal{v} = 0$, and tune the pattern's shifts by choosing vortex
positions in the range $0.3 (\to \textrm{large shifts}) < |x_\textnormal{v}|/W
< 0.8 (\to \textrm{small shifts})$, see Fig.~1C (detailed parameters are given
in the Supplementary Text).  Note that in Figs.~1, E and
F, vortices have left the device at zero field $B = 0$ in both cases. This is
different in Fig.~3 where the central peak appears shifted in some of the
traces due to the presence of vortices remaining at vanishing field. Pushing
the field amplitude $|B|$ too far up, beyond $\SI{10}{mT}$, the patterns turn
blurred and the observation of individual jumps gets hampered.

A further test of our interpretation of pattern-shifts due to vortex jumps is
provided by the minimal field $B_\textnormal{m} = \pi\Phi_0/4 W^2$ for
metastable vortex configurations, see Ref.~\cite{Gurevich1994} and the
Supplementary Text.  Using the results in Ref.\ \cite{Gurevich1994}, we find
the first (meta-)stable position inside the film appearing at $\Phi_W
(B_\textnormal{m})/\Phi_0 \approx 0.8$, i.e., within the main dome
($|\Phi_W|/\Phi_0 < 1.4$) of the Fraunhofer pattern.  Typical barriers the
vortices have to overcome are of order \cite{Gurevich1994} $\varepsilon_0 d$,
see fig.~S18A in the Supplementary Text, where $\varepsilon_0 = (4\pi/\mu_0)
(\Phi_0/4\pi\lambda_\textnormal{L})^2$ is the vortex line energy.  On the
other hand, analyzing the time traces of fluctuating currents, we can derive
an energy scale for the typical barriers $U_\textnormal{s}$ for vortex entry:
given a barrier $U_\textnormal{s}$ at temperature $T$ and an attempt time
$t_0$, a vortex requires a time $t \sim t_0
\exp[U_\textnormal{s}/T]$ to overcome the barrier (we set
$k_\textnormal{B}$ to unity).  Knowing the typical dynamical time $t$ for
shifts in the Fraunhofer pattern, we can derive a quite accurate estimate for
the barrier $U_\textnormal{s} \approx T \ln(t/t_0) \approx
\SI{1.8}{K}$ (we choose $T = \SI{55}{mK}$, an attempt time
\cite{Malozemoff1990, Koshelev1990} $t_0 \sim \SI{e-11}{s}$, and a typical
dwell time $t \sim \SI{e3}{s}$ as observed in our experiments, see fig.~S14).
Equating the two, $U_\textnormal{s} \approx \varepsilon_0 d$, we extract
$\lambda_\textnormal{L} \sim \SI{3.3}{\upmu m}$ for the London length. These
results are in good accord with the recent determination of the superfluid
stiffness $\rho_\textnormal{s}$ via kinetic inductance measurements in twisted
films, see Refs.~\cite{Kim2024,Oliver2024}. Indeed, the reported
\cite{Kim2024} value $\rho_\textnormal{s} \approx {0.4-0.5}\,\mathrm{K}$
around optimal gating is quite consistent with our estimate for the
fluctuation barrier $\varepsilon_0 d \approx (4/3) \pi \rho_\textnormal{s}
\approx 1.6 - 2\,\mathrm{K}$ (where the factor $4/3$ accounts for the
difference in film thicknesses).

\paragraph*{Vortex fluctuations} We close our discussion with the analysis of
another data set with the leads tuned to the edge of the superconducting
region, see purple rhombus and red dot in Fig.~2A ($n_\textnormal{l} =
\SI{4.8e12}{cm^{2}}$, $D_\textnormal{l}/\epsilon_0 = \SI{-0.3}{V/nm}$ and
$n_\textnormal{j} = \SI{6.2e12}{cm^{-2}}$, $D_\textnormal{j}/\epsilon_0 =
\SI{-0.5}{V/nm}$). This choice of parameters leads to a reduction in the
superfluid stiffness in the lead, see Ref.\ \cite{Kim2024}.  The Fraunhofer
pattern in Fig.~4A exhibits pronounced bi-stability effects in
$I_\textnormal{cj}(B)$.  These derive from rapid switchings between
superconducting (purple circle) and dissipative (orange square) states in the
$V$--$I$ traces as shown in Fig.~4B and its inset (recorded at $B =
\SI{1.8}{mT}$, black arrows in Fig.~4A).  The timescale of these fluctuations
is on the order of seconds, see the time-dependent measurement in Fig.~4D
taken at $I=\SI{2.5}{nA}$ (red dashed line in Fig.~4B).  The counting
statistics of these fluctuations is analyzed in the Supplementary Text, see
fig.~S8.

We attribute these rapid switchings between states to vortex
fluctuations.  As seen before, a vortex jumping into and out of the leads
produces shifts in the Fraunhofer pattern. Fixing the field $B^\ast$ and the
bias current $I^\ast$ at a point where the presence (absence) of a vortex
causes $I_\textnormal{cj}$ to be lower (higher) than the bias current
$I^\ast$, see Fig.~4C, produces the observed jumps in voltage traces which we
correlate with jumps of vortices. The reduced superconducting stiffness
results in a smaller barrier $U_\textnormal{s} \approx \varepsilon_0 d$ for
thermal vortex dynamics and thus a shorter dwell time $t$ between jumps.
Indeed, the typical dwell time in Fig.~4D is of order $t \sim \SI{1}{s}$ and
following the argument presented above, we find that
$U_\textnormal{s} \approx T \log (t/t_0) \approx \SI{1.4}{K}$,
about 3/4 of the value found previously.

%Conclusion
%
\paragraph*{Conclusion} In conclusion, we have demonstrated the formation of a
gate-defined Josephson junction in MAT4G and its application as a vortex
sensor. The measured interference patterns agree excellently with the
predicted behaviour for a weak transverse screener with a Pearl length
surpassing the sample dimension, $\Lambda > W$. Our data enable us to
precisely track the entry and exit of vortices. By fitting the data, we can
accurately position the vortices within the leads. Observing vortices in MAT4G
opens new avenues for exploring vortex motion in this new class of twisted
materials. Gaining deeper insights into their dynamics offers a significant
advance in the engineering of thin-film superconducting devices.

\bibliography{scifile24Sept}

\bibliographystyle{Science}

\newpage

\begin{figure}
%\centering
\includegraphics[width=1\textwidth]{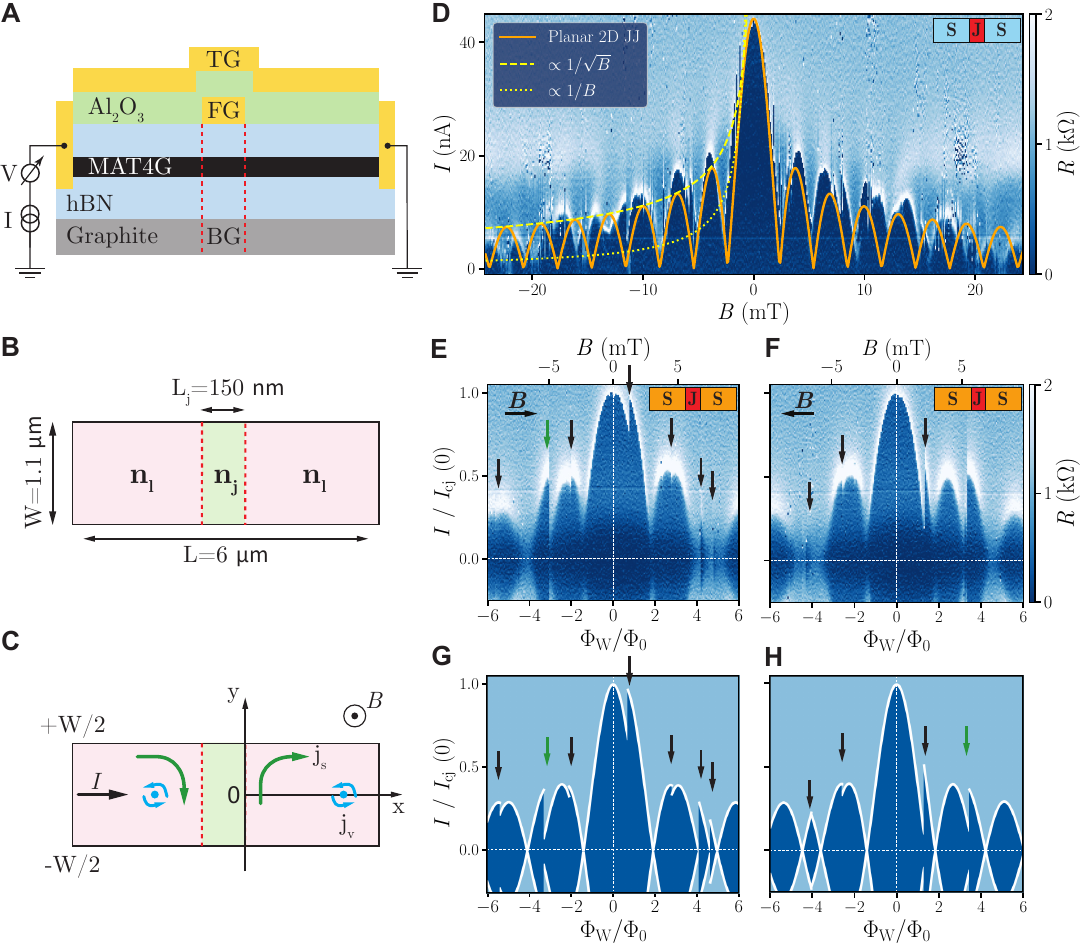}
\caption{\textbf{Josephson junction in MAT4G.} (A) Schematic cross-section of
the JJ and measurement setup, where $V$ is the measured
voltage drop and $I$ is the applied current. (B) Schematic top view of the film with a JJ of width
$W$, length $L_\textnormal{j}$ and overall length $L$. The densities in the leads and junction are denoted by
$n_\textnormal{l}$ and $n_\textnormal{j}$. (C) Schematic configuration with
two vortices placed in the leads.  The magnetic field $B$ is perpendicular to
the sample and the transport current $I$; counterrotating screening and vortex
current densities $j_\textnormal{s}$ and $j_\textnormal{v}$ are indicated. (D)
Differential resistance $R$ measured as a function of $I$ and $B$ for
$n_\textnormal{l} = \SI{-3.5e12}{cm^{-2}}$, $D_\textnormal{l}/\epsilon_0 =
\SI{0.2}{V/nm}$ and $n_\textnormal{j} = \SI{-6.3e12}{cm^{-2}}$,
$D_\textnormal{j} / \epsilon_0 = \SI{0.45}{V/nm}$ (BRB setting in Fig.~2A). The orange and yellow dashed lines show the
theoretical predictions for a 2D JJ under weak screening conditions, whereas the dotted yellow line shows the rapid decay $\propto 1/B$ for a standard JJ . (E and F) Differential resistance $R$ measured as a function of $I$ and $B$, as well as flux $\Phi_W/\Phi_{0}$, with the flux $\Phi_W = B W^2$ penetrating
junction and leads, for a forward magnetic field sweep in (E) and a backward
sweep in (F).  The tuning parameters are $n_\textnormal{l} =
\SI{4.2e12}{cm^{-2}}$, $D_\textnormal{l}/\epsilon_0 = \SI{-0.3}{V/nm}$ and
$n_\textnormal{j} = \SI{6.2e12}{cm^{-2}}$, $D_\textnormal{j}/\epsilon_0 =
\SI{-0.5}{V/nm}$ (ORO setting in Fig.~2A).
Sudden shifts of the interference pattern are highlighted with black arrows.
The green arrows highlight two shifts which are highly reproducible between
measurements and are symmetric in $B$. (G) and (H) Theoretical fits of (E)
and (F) on the basis of Ref.~\cite{KoganMints2014}.}
\label{fig:1}
\end{figure}

\begin{figure}[t!]
%\centering
\includegraphics[width=1\textwidth]{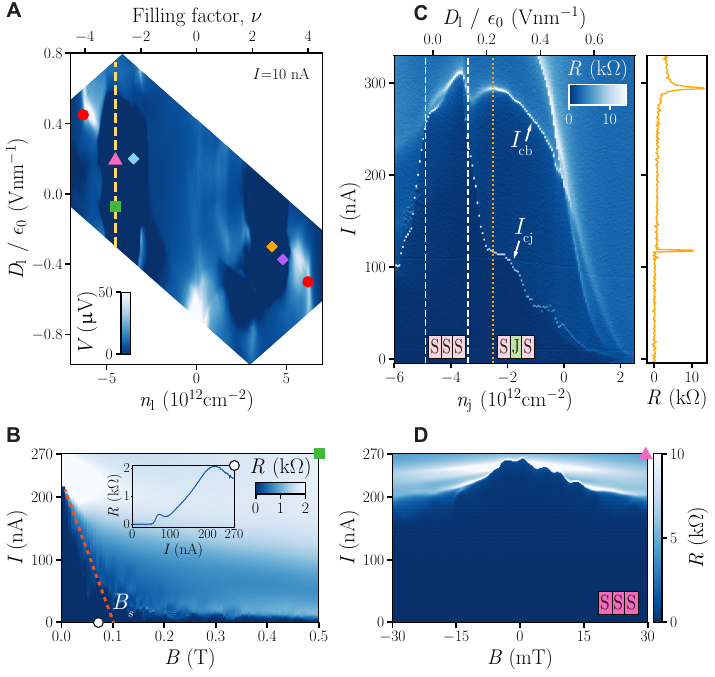}
\caption{\textbf{Phase diagram and junction tuning.} (A) Voltage $V$ versus
leads' density $n_\textnormal{l}$ and displacement field $D_\textnormal{l}$,
measured in a 4-terminal configuration not crossing the junction at a constant
current $I=\SI{10}{nA}$. The filling factor $\nu$ is plotted on the top axis.
The blue dark regions signal superconductivity. Red dots indicate full filling
where the junction is tuned into the resistive state. (B) Differential
resistance $R$ measured as a function of $I$ and $B$ with the same setup as in
(A) (device tuned to the green square in (A)). The orange dashed line
indicates the fit to extract the surface penetration field $B_\textnormal{s}$,
with the inset showing a line trace recorded at the white circle
$B=\SI{72}{mT}$.  (C) Differential resistance $R$ measured as a function of
$I$ while sweeping $n_\textnormal{j}$ and keeping $n_\textnormal{l}$ fixed.
$D_\textnormal{j}$ is fixed at zero whereas $D_\textnormal{l}$ is swept as
shown on the top axis, following the yellow dashed line in (A). By sweeping
$n_\textnormal{j}$, the junction can be tuned from a resistive to a
superconducting state, resulting in a SJS or SSS configuration, with the
critical currents of the bulk and junction denoted by $I_\textnormal{cb}$ and
$I_\textnormal{cj}$. A line trace of the differential resistance $R$ is shown
with $n_\textnormal{j}$ fixed at the dotted orange line. (D) Differential
resistance $R$ measured as a function of $I$ and $B$ with the entire device
tuned to the pink triangle shown in (A), implying that no junction is formed.
This SSS configuration exhibits no interference pattern.} \label{fig:2}
\end{figure}

\begin{figure}[t!] %\centering
\includegraphics[width=0.5\textwidth]{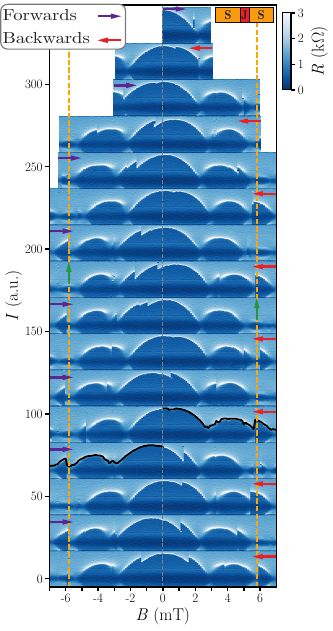}
\caption{\textbf{Vortex jumps.} (A) Differential resistance $R$ measured as a
function of $I$ (in arbitrary units) and $B$ for the same tuning as in
Figs.~1, E and F (ORO setting). Measurements with increasing field sweeps are
recorded from top to bottom. Purple arrows indicate a forward sweep in $B$,
whereas red arrows indicate a backward sweep.  Between each measurement, the
device is kept in the same configuration.  The orange dashed lines indicate
the position at which reproducible shifts are observed, with two highlighted
with green arrows (same jumps as in Figs.~1, E and F). Black traces emphasize
the symmetry between forward/backward sweeps.} \label{fig:3}
\end{figure}

\begin{figure}[t!]
%\centering
\includegraphics[width=1\textwidth]{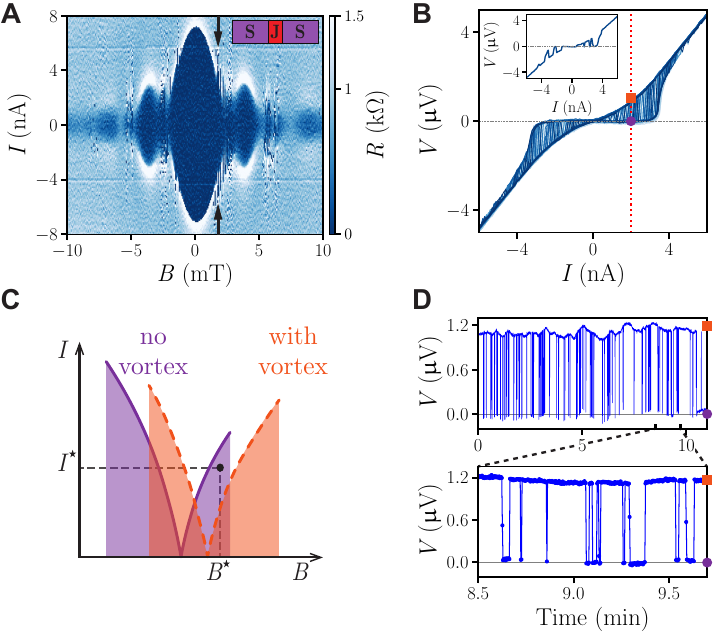}
\caption{\textbf{Vortex fluctuations.} (A) Differential resistance $R$
measured as a function of $I$ and $B$ with parameters $n_\textnormal{l} =
\SI{4.8e12}{cm^{-2}}$, $D_\textnormal{l}/\epsilon_0 = \SI{-0.3}{V/nm}$ and
${n_\textnormal{j} = \SI{6.2e12}{cm}^{-2}}$, $D_\textnormal{j}/\epsilon_0 =
\SI{-0.5}{V/nm}$, see purple rhombus and red dot in Fig.~2A.  Pronounced
fluctuations in the critical current are visible in the interference pattern.
(B) Seventy current--voltage traces measured consecutively while keeping $B =
\SI{1.8}{mT}$ fixed at the position indicated by the black arrows in (A). The
various traces show bistable fluctuations in the measured voltage. The inset
shows one single time trace, where the switching between resistive (normal
state, orange square) and zero voltage (superconducting state, purple circle)
is observed. (C) Illustration of a Fraunhofer pattern without (purple) and
with (orange) a vortex; vortex jumps at fixed $I^\star$ and
$B^\star$ trigger transitions between superconducting (purple) and
normal (orange) states. (D) Voltage as a function of time at a fixed applied
current $I = \SI{2.5}{nA}$ marked by the red dotted line in (B).  The voltage
jumps between zero voltage (superconducting state, purple circle) and
$\SI{1.2}{\upmu V}$ (normal state, orange square).} \label{fig:4}
\end{figure}

\clearpage

\section*{Acknowledgments} We thank Peter M\"{a}rki and the staff of the ETH
cleanroom facility FIRST for technical support. We acknowledge fruitful
discussions with Vladimir Kogan and Manfred Sigrist. We thank Giulia Zheng for
her support during the project. Financial support was provided by the European
Graphene Flagship Core3 Project, H2020 European Research Council (ERC) Synergy
Grant under Grant Agreement 951541, the European Union’s Horizon 2020 research
and innovation program under grant agreement number 862660/QUANTUM E LEAPS,
the European Innovation Council under grant agreement number
101046231/FantastiCOF, the EU Cost Action CA21144 (SUPERQUMAP), and NCCR QSIT
(Swiss National Science Foundation, grant number 51NF40-185902).  K.W. and
T.T. acknowledge support from the JSPS KAKENHI (Grant Numbers 21H05233 and
23H02052) and the World Premier International Research Center Initiative
(WPI), MEXT, Japan. F.G. is grateful for the financial support from the Swiss
National Science Foundation (Postdoc.Mobility Grant No. 222230).

\paragraph*{Author contributions:} M.P. fabricated the device. T.T. and K.W.
supplied the hBN crystals. M.P. and C.G.A. performed the measurements. M.P.
and C.G.A. analyzed the data. F.G., V.G. and G.B. developed the theoretical
model. M.P. and G.B. wrote the manuscript, and all authors were involved in
the reviewing process. M.P., E.P., K.E. and T.I. conceived and designed the
experiment. T.I. and K.E. supervised the work.

\paragraph*{Competing interests:}
The authors have no competing interests.

\paragraph*{Data and materials availability:} The data that support the
findings of this study will be made available online through the ETH Research
Collection.

%Here you should list the contents of your Supplementary Materials -- below is
%an example.  You should include a list of Supplementary figures, Tables, and
%any references that appear only in the SM.  %Note that the reference numbering
%continues from the main text to the SM.  % In the example below, Refs. 4-10
%were cited only in the SM.

% For your review copy (i.e., the file you initially send in for
% evaluation), you can use the {figure} environment and the
% \includegraphics command to stream your figures into the text, placing
% all figures at the end.  For the final, revised manuscript for
% acceptance and production, however, PostScript or other graphics
% should not be streamed into your compliled file.  Instead, set
% captions as simple paragraphs (with a \noindent tag), setting them
% off from the rest of the text with a \clearpage as shown  below, and
% submit figures as separate files according to the Art Department's
% instructions.

\clearpage

\end{document}